# Proton induced reactions on $^{natural}$U at 62.9 MeV

**A. Guertin, S. Auduc, G. Rivière, P. Eudes, F. Haddad, C. Lebrun, T. Kirchner**
*SUBATECH, F- 44307 Nantes cedex 03, France*
**C. Le Brun, F.R. Lecolley, J.F. Lecolley, M. Louvel, F. Lefebvres, N. Marie, C. Varignon**
*Laboratoire de Physique Corpusculaire, F-14050 Caen cedex, France*
**X. Ledoux, Y. Patin, Ph. Pras**
*C.E.A. Bruyere-le-Chatel, France*
**Th. Delbar, A. Ninane**
*Institut de Physique Nucléaire, 1348 Louvain-la-Neuve, Belgium*
**L. Stuttge**
*Institut de Recherches Subatomiques, F-67037, STRASBOURG cedex 2, France*
**F. Hanappe**
*U.L.B., Bruxelles, Belgium*

## Introduction

Double differential cross sections (DDCS) for light charged particles (proton, deuteron, triton, $^3$He, alpha) and neutrons produced by a proton beam impinging on a $^{238}$U target at 62.9 MeV were measured at the CYCLONE facility in Louvain-la-Neuve (Belgium). These measurements have been performed using two independent experimental set-ups ensuring neutron (DeMoN counters) and light charged particles (Si-Si-CsI telescopes) detection. The charged particle data were measured at 11 different angular positions from 25° to 140° allowing the determination of angle differential, energy differential and total production cross sections.

## Experimental set-up

The experiment intended to measure at the same time neutrons and light charged particles. Two independent set-ups have been used to achieve this goal.

The neutrons were detected using five DeMoN counters [Til95], large volume NE213 liquid scintillator, placed respectively at 24°, 35°, 55°, 80° and 120° [LPC01]. A lead cylinder installed inside a "bombarde" barrel filled with paraffin and boron - materials that are efficient shields against low-energy background neutrons - surrounded each detector resulting in a signal-to-noise ratio of 2 to 1. A very good discrimination between neutrons and gamma rays is achieved using pulse shape analysis and neutron energies were derived from the time-of-flight method, taking into account the depth at which the particle interacts inside the detector [Lec96]. These data are under analysis by the LPC Caen.

Light charged particles have been detected using a set of five triple telescopes (Si-Si-CsI) settled inside a vacuum chamber. A schematic view of a telescope is presented on figure 1.

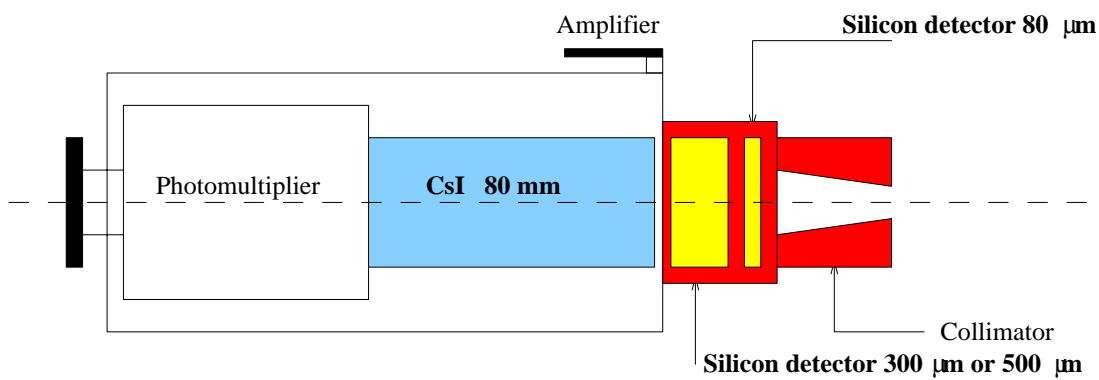

Figure 1: Schematic view of the triple telescope (Si-Si-CsI).

Each telescope consists of two silicon detectors (80 µm thick (Si1) and 300 or 500 µm thick (Si2) respectively), followed by a one inch diameter and 80 mm thick CsI(Tl) crystal. It allows us to identify and to measure the charged particles over their entire energy range. A copper collimator is placed in front of each telescope to precisely define the detection solid angle. The thickness of the first silicon detector determines the low energy thresholds of our set-up, which are respectively 2.5 MeV for protons, 3 MeV for deuterons, 3.5 for tritons, 9 MeV for $^3$He and 10 MeV for alpha particles.

Several different targets were used during the experiment. An enriched $^{238}$U target of 47.6 mg/cm$^2$ was used for the measurement. The calibration of the detectors was performed using a $(C_3H_6)_n$ (4.47 mg/cm$^2$) and a $^{12}$C (8.96 mg/cm$^2$) targets.

A group of two telescopes were mobile and allowed us to record data from 25° up to 95° by step of 10° (forward angles).
During all the experiment, two telescopes were set at backward angles in order to accumulate sufficient statistics for 110°, 130° and 140°. The last one was set at 30° to be used as a second beam monitor. A Faraday cup placed at the end of the beam line recorded the integrated beam current.

Data Analysis

Energy calibration and particle identification

The energy calibration of the CsI has been made for each type of particle. It is based on the kinematics of the elastic scattering reaction on $(C_3H_6)_n$ target and nuclear reactions. The H(p,p) reaction allows by changing the angular position of the detectors, to get outgoing protons with energy from 51.7 MeV at 25° to 11.2 MeV at 65°. Peaks for composite particles are coming from nuclear reactions on $^{12}$C. To get the silicon calibration curves, in addition to an alpha source (Pu, Am and Cu), we use the H(p,p) reaction at large angle for which the energy deposition is large on the Si2 detector.

The identification of the charged particles is obtained using the well-known ΔE-E method. The two first stages of the telescopes (Si1-Si2) are used for the slowest particles, while the two last stages (Si2-CsI) take over for the most energetic particles. In addition, most of the background coming from gamma is suppressed using a pulse shape analysis of the CsI energy signal. Unambiguous particle identification can be obtained over the entire energy range by combining the information contained in these different plots.

Normalization of the absolute cross sections has been derived from the Faraday cup information. The statistical errors correspond to 10% for 10 µb/(sr.MeV). Some corrections must be applied to our spectra. Indeed, some particles may have scattered on the collimator before being detected. Slight distortions of the spectra are thus expected and have to be corrected. An estimation of this effect was done using simulations based on the GEANT code [Geant], which allows taking into account the geometry of the experimental set-up. We found that up to 5% of the measured events per MeV have to be removed from the spectra for protons, this correction becoming negligible in the case of heavier particles. In addition to this effect, nuclear reactions may occur in the CsI, which lead to a bad measurement of the particle energy. This effect is assumed to be small at this energy.

## Results

DDCS for protons, deuterons, tritons, $^3$He and alpha are available at 11 angular positions from 25° to 140°. In this paper, we will focus on proton spectra. These data are reported on figure 2 for all angles.

At all angles, a contribution from elastic scattering of protons on lead is present in the high-energy part of the spectra (above 55 MeV). At backward angles ($\theta \geq 130°$), the spectra exhibit a characteristic evaporation behavior with approximately the same slope. Below 130°, energy spectra show a strong angular dependence characteristic of pre-equilibrium processes. At the most forward angles, energy spectra are totally dominated by pre-equilibrium effects and are almost flat.

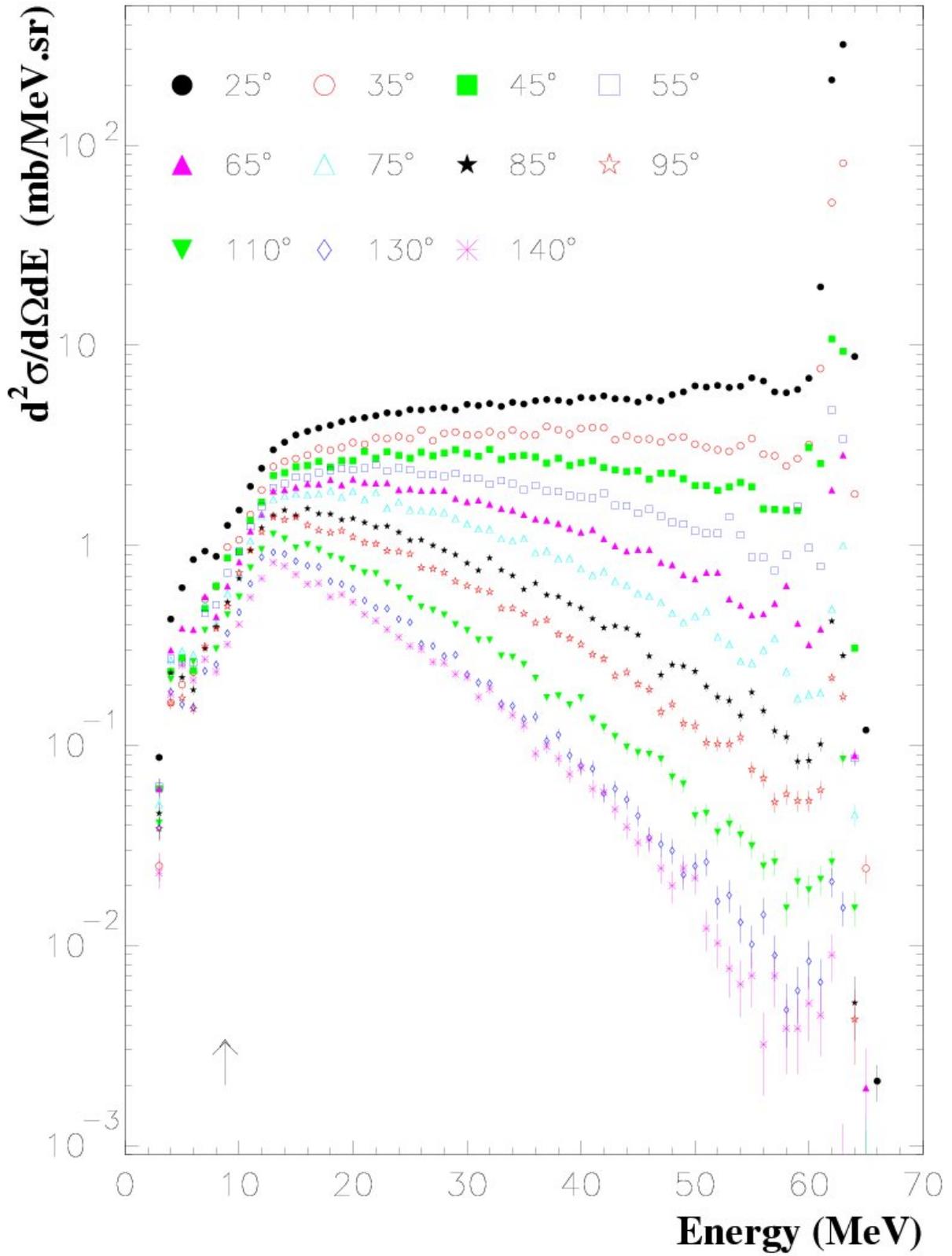

Figure 2: $^{238}$U(p, xp) double differential production cross section at 62.9 MeV.

The data analysis is still in progress, it will be possible soon to extract the angular and the energy differential cross sections from our measurements. The angular differential cross sections dσ/dΩ were obtained by integrating over the energy the double differential cross section whereas the energy differential cross sections dσ/dE are derived directly by fitting our data points using the Kalbach systematic [Kal81],[Kal88].

Finally, the production cross sections will be obtained for light charged particles by integrating over the energy the energy differential cross sections dσ/dE.

## Comparisons with theoretical calculations

The DDCS measurements present a very sensitive and difficult challenge for the theoretical models. Within the HINDAS collaboration, a new nuclear model code TALYS is being developed for the 20 MeV - 200 MeV energy range. It results of collaboration between NRG-Petten (Netherlands) and CEA-Bruyères-le-Châtel (France). This global program intends to describe not only total and partial cross sections but also the energy spectra, angular distributions and double differential cross sections of the neutrons, photons, light charged particles and residuals emitted in the nuclear reactions. In addition, to improve optical model and direct interaction models, special attention has been paid to the description of the pre-equilibrium processes, where the two-component exciton model has been extended to an arbitrary number of reaction steps [Talys]. TALYS calculations of double differential cross sections on uranium should be performed within a year. A close collaboration with the theoreticians developing the TALYS code has shown fruitful results (lead target) [Gue02] in the quality of the model predictions.

## Conclusion

Light charged particle double differential cross sections have been measured in 62.9 MeV proton-induced reactions on $^{238}$U target. Values have been extracted over 11 angular positions from 25° to 140°. Other targets (Co) and incident energies (135 MeV) are still under analyses.

This work has been supported partly by the European Communities under contract number FIKW-CT-20000-00031 by the GDR GEDEON (research group CEA - CNRS - EDF - FRAMATOME).}